\documentclass[shortnote,twocolumn,letterpaper]{jpsj3}
\usepackage{txfonts}
\usepackage{amsmath,amsfonts,amssymb}
\usepackage[usenames]{color}

\title{An inequality for spinor Bose-Einstein condensates}
\author{Daisuke A. Takahashi$^{1,2}$\thanks{daisuke.takahashi.ss@riken.jp}}
\inst{$^1$RIKEN Center for Emergent Matter Science (CEMS), Wako, Saitama 351-0198, Japan \\
$^2$Research and Education Center for Natural Sciences, Keio University, Hiyoshi 4-1-1, Yokohama, Kanagawa 223-8521, Japan}

\abst{
 An inequality for spin-$F$ Bose-Einstein condensates (BECs) $ F^2(\rho^2-|\Theta|^2)-\boldsymbol{M}^2\ge0 $ is reported, where $\rho$, $\Theta$, and $\boldsymbol{M}$ represent the density, singlet pair amplitude, and magnetization vector, respectively. The distribution of high-symmetry spinors in the allowed region by the inequality is elucidated with using the Majorana representation. The result is illustrated by the example of spin-2 BECs.
}

\begin{document}
\maketitle
\begin{figure}[t]
		\begin{center}
		\includegraphics{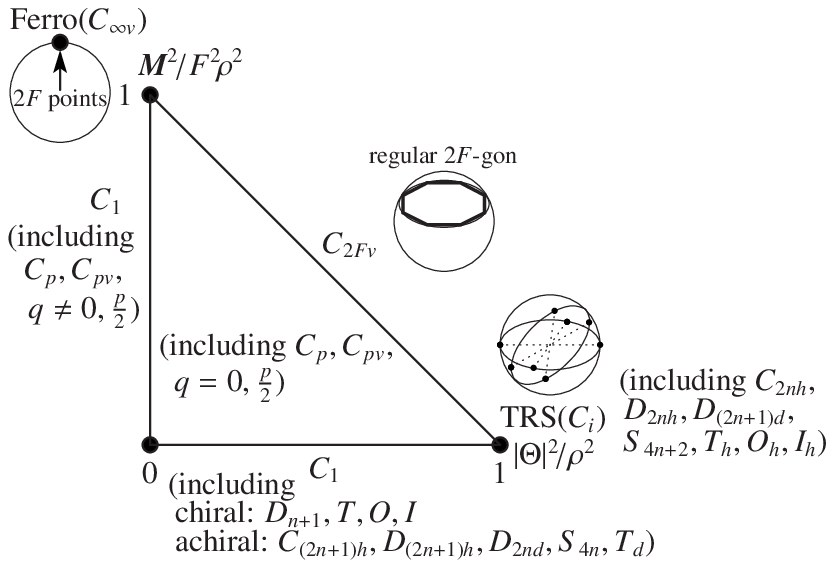}
		\caption{\label{fig:spinFtriangle}The triangle region where spinors inhabit. The MRs of ferromagnetic states have degenerate $ 2F $ points, and those of TRS states have the inversion symmetry. These two states are connected via  $ C_{2Fv} $-symmetric states, having the regular $ 2F $-gon MR. If we continuously change the spinor along the line $ \boldsymbol{M}^2/F^2=\rho^2-|\Theta|^2 $, the spinor becomes the $ D_{2Fh} $-symmetric state $ \boldsymbol{\psi}=\sqrt{\rho/2}(1,0,\dots,0,1)^T $  at the TRS point, represented by the regular $ 2F $-gon on a great circle. The TRS point, however, contains more diverse states. TRS states may include higher-symmetry states $ C_{2nh}, D_{2nh}, D_{(2n+1)d}, S_{4n+2}, T_h, O_h, I_h \ (n\ge 1)$, and the states with these symmetries can appear only at this point. Similarly, the states with the symmetry $ D_{n+1},T,O,I,C_{(2n+1)h},D_{(2n+1)h},D_{2nd},S_{4n},T_d \  (n\ge 1)$ can appear only on the edge $ \boldsymbol{M}^2=0 \ \& \ 0\le|\Theta|^2<\rho^2 $. Here, the term chiral (achiral) means that there exists no (at least one) improper rotation. The line $ |\Theta|^2=0 $ includes  $ C_{p} $-symmetric states defined as follows \cite{PhysRevA.85.051606}. Let $ p,q $ be integers s.t.  $ 0<2q<p<2F $. Then, the spinors satisfying ``$ \psi_m\ne0 $  $ \rightarrow $  $ m\equiv q \text{ mod } p $'' has the $ C_p $-symmetry and $ |\Theta|=0 $. If the relative phases of $ \psi_m $'s are adjusted, the symmetry becomes $ C_{pv} $. The examples are the H, I, J phases in spin-3 BECs\cite{PhysRevLett.96.190404,PhysRevLett.96.190405,PhysRevA.76.013605,PhysRevA.84.053616}. When $ q=0 $ or $ \frac{p}{2} $, the symmetry group is the same but $ |\Theta|\ne 0 $, so the spinors are not on the edge but in the triangle.}
		\end{center}
	\end{figure}
\begin{figure}[t]
		\begin{center}
		\includegraphics{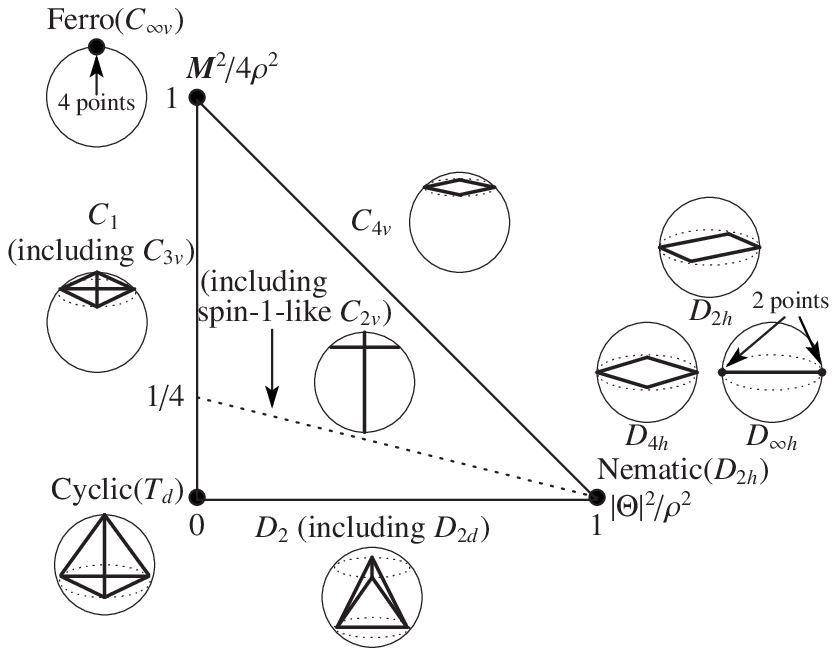}
		\caption{\label{fig:spin2region} The triangle region for spin-2 BECs. Three vertices represent ferromagnetic, nematic, and cyclic states. Here we provide expressions of spinors using $ \boldsymbol{\psi} $ or $ T $ [Eq.~(\ref{eq:spin2tracelesstensor})]. Ferro($C_{\infty v}$): $ \boldsymbol{\psi}\propto(1,0,0,0,0)^T $. Cyclic($T_d$): $ T\propto\operatorname{diag}(\mathrm{e}^{2\pi\mathrm{i}/3},\mathrm{e}^{4\pi\mathrm{i}/3},1) $. Nematic($D_{2h}$): $ T\propto\operatorname{diag}(\cos(\eta-\frac{2\pi}{3}),\cos(\eta+\frac{2\pi}{3}),\cos\eta),\, \eta\in \mathbb{R} $.   $ D_2 $:  $ T=\operatorname{diag}(\alpha_1,\alpha_2,-\alpha_1-\alpha_2), \alpha_i\in\mathbb{C} $.  $ D_{2d} $: $ \alpha_2=\alpha_1^* $ in $ D_2 $.  $ C_{4v} $:  $ \boldsymbol{\psi}=(\psi_2,0,0,0,\psi_{-2})^T $.   $ C_{3v} $:  $ \boldsymbol{\psi}=(\psi_2,0,0,\psi_{-1},0)^T $. Spin-1-like $ C_{2v} $: $ \boldsymbol{\psi}=(0,\psi_1,0,\psi_{-1},0)^T $. In nematic states, higher-symmetry states called the uniaxial ($D_{\infty h},\ \eta=0$) and biaxial ($D_{4h},\ \eta=\frac{\pi}{2}$) nematic states exist, which are favored in the presence of quantum correction \cite{PhysRevLett.98.160408,PhysRevLett.98.190404,PhysRevA.81.063632}. The cyclic state has both $ C_{3v} $ and $ D_{2d} $ symmetries. Note that the uniqueness of the spinor under the condition $ |\Theta|^2=\boldsymbol{M}^2=0 $ is specific to spin-2.}
		\end{center}
	\end{figure}

	Spinor Bose-Einstein condensates (BECs) in ultra cold atomic gases offer opportunities to investigate rich phases and topological defects emerging due to multicomponent nature of order parameters \cite{Kawaguchi:2012ii,RevModPhys.85.1191}. The symmetry of spinors can be well understood by visualization such as spherical harmonics plots \cite{Kawaguchi:2012ii} or the Majorana representation (MR) \cite{PhysRevLett.97.180412,PhysRevA.84.053616,PhysRevA.85.051606}. However, they cannot provide an intuition for energetic properties, which are essential for determination of ground-state phase diagram and stable profiles of cores of topological defects \cite{Kobayashi09}. 
To understand the energetics, inequalities between  $ U(1)\times SO(3) $-invariant scalars are important, since the interaction terms in the Hamiltonians are written by them. 
	In this short note we report one such inequality for spinor BECs applicable for arbitrary integer spins, and reveal the distribution of high-symmetry spinors using the MR. \\
	\indent Let $ \boldsymbol{\psi}=(\psi_F,\dots,\psi_{-F})^T $ be an order parameter of the spin-$ F $ BEC. Then, the main result of this note is given by
	\begin{align}
		\rho^2-|\Theta|^2-\frac{\boldsymbol{M}^2}{F^2}\ge0, \label{eq:mainclaim}
	\end{align}
	where  $ \rho=\sum_{m=-F}^F|\psi_m|^2 $ is a density, $ \Theta=\sum_{m=-F}^F(-1)^m\psi_m\psi_{-m} $ is a singlet pair amplitude, and $ \boldsymbol{M}=(M_x,M_y,M_z)^T,\ M_i=\boldsymbol{\psi}^\dagger S_i\boldsymbol{\psi}, $ is a magnetization vector with $ S_i $'s being spin-$ F $ matrices. 
	Note that $ \Theta $ is the inner product between $ \boldsymbol{\psi} $ and its time-reversed state $ \mathcal{T}\boldsymbol{\psi}$, where we define $ [\mathcal{T}\boldsymbol{\psi}]_m:=(-1)^m\psi_{-m}^* $. 
	The proof of Eq.~(\ref{eq:mainclaim}) is given in the final part of this note. \\
	\indent The inequality (\ref{eq:mainclaim}) can be regarded as a refinement of the well-known inequality $ |\Theta|^2\le\rho^2 $, and provides a more precise upper bound for the magnitude of magnetization depending on the value of $ |\Theta|^2 $. The equality of Eq.~(\ref{eq:mainclaim}) occurs in the following two cases: (i) The time-reversal-symmetric (TRS) states ($ \boldsymbol{\psi}\propto \mathcal{T}\boldsymbol{\psi} $), in which $ \rho=|\Theta| $ holds and the magnetization vanishes, and (ii) $ C_{2Fv} $-symmetric states $ \boldsymbol{\psi}=(\psi_F,0,\dots,0,\psi_{-F}) $.  The inequality (\ref{eq:mainclaim}) holds not only for spatially-uniform states but also for every point in nonuniform states such as textures or vortices. \\
	\indent Combining the inequality (\ref{eq:mainclaim}) with the obvious inequalities $ \boldsymbol{M}^2\ge0 $ and $ |\Theta|^2\ge0 $, we can conclude that any spinor lies in the triangle region shown in Fig.~\ref{fig:spinFtriangle}. Let us discuss the symmetries and the MRs in Fig.~\ref{fig:spinFtriangle}. Henceforth, we use the standard Schoenflies notation to describe the point groups. The MR is introduced as follows \cite{PhysRevLett.97.180412,PhysRevA.85.051606,PhysRevA.84.053616}. Let us consider the following equation for spin-$F$ spinor $ \boldsymbol{\psi}=(\psi_F,\dots,\psi_{-F})^T $:
	\begin{align}
		\sum_{m=-F}^F\psi_m(-\zeta)^{F+m}\sqrt{\binom{2F}{F+m}}=0. \label{eq:majorana}
	\end{align}
	Then, parametrizing  $ 2F $ roots of Eq. (\ref{eq:majorana}) by $ \zeta=\mathrm{e}^{\mathrm{i}\varphi}\tan\frac{\theta}{2} $, and plotting $ 2F $ points $ (\theta,\varphi) $ on the unit sphere, we obtain the MR for a given spinor. If $ \psi_F=\dots=\psi_{F-n+1}=0 $, we must put $ n $ points at the south pole, corresponding to the solution $ \zeta=\infty $. 
	The MRs of ferromagnetic states are given by degenerate $ 2F $ points at the same position, having a $ C_{\infty v} $-symmetry.  
	The MRs of TRS states are characterized by $ C_i $-symmetry, because the time-reversal transformation is represented by the inversion operation. 
	Some particular TRS states may have higher symmetries (See Fig.~\ref{fig:spinFtriangle}). 
	On the edge of the triangle $ \boldsymbol{M}^2/F^2=\rho^2-|\Theta|^2 $ with $ 0< |\Theta|<\rho $, the state is given by $ C_{2Fv} $-symmetric $ \psi=(\psi_F,0,\dots,0,\psi_{-F})^T $, whose MR is given by a regular $ 2F $-gon. 
	The other edges, $ \boldsymbol{M}^2=0 $ and $ |\Theta|^2=0 $, are not characterized by one group symmetry, but several important high-symmetry states exist on these edges (See Fig.~\ref{fig:spinFtriangle} and its caption). 
	The remaining degrees of freedom for each point in the triangle are not the same; while the ferromagnetic state is unique up to $ U(1)\times SO(3) $ transformation and no additional parameter remains, the set of TRS states has $ 2F-3 $ parameters if $ F\ge 2 $. \\
	\indent Let us see the examples for small spins. For spin-1, the identity $ \rho^2-|\Theta|^2=\boldsymbol{M}^2 $ holds and hence $ \boldsymbol{M}^2 $ and $ |\Theta|^2 $ are not independent, and the triangle of Fig.~\ref{fig:spinFtriangle} shrinks to one line. Indeed, any spin-1 state can be transformed to $ \boldsymbol{\psi}=\tfrac{1}{2}(\sqrt{\rho+|\Theta|}+\sqrt{\rho-|\Theta|},0,\sqrt{\rho+|\Theta|}-\sqrt{\rho-|\Theta|})^T $ by a $ U(1)\times SO(3) $ transformation. The state with $ 0<|\Theta|<\rho $  has a $ C_{2v} $-symmetry. The ground state is realized by maximizing (minimizing) the value of $ |\Theta|^2 $, and the state becomes the polar (ferromagnetic) state \cite{JPSJ.67.1822,Ho:1998zz}, having the symmetry $ D_{\infty h} $ ($ C_{\infty v} $). \\
	\indent Next, let us consider the spin-2 BECs. The triangle region and several important states with their MRs are given by Fig.~\ref{fig:spin2region}. 
	It is convenient to introduce the traceless symmetric tensor \cite{PhysRevA.61.033607,Kawaguchi:2012ii}
	\begin{align}
		T= \begin{pmatrix} \frac{\psi_2+\psi_{-2}}{2}-\frac{\psi_0}{\sqrt{6}} & \frac{\mathrm{i}(\psi_2-\psi_{-2})}{2} & \frac{\psi_{-1}-\psi_1}{2} \\ \frac{\mathrm{i}(\psi_2-\psi_{-2})}{2} & -\frac{\psi_2+\psi_{-2}}{2}-\frac{\psi_0}{\sqrt{6}} & -\frac{\mathrm{i}(\psi_1+\psi_{-1})}{2} \\ \frac{\psi_{-1}-\psi_1}{2} & -\frac{\mathrm{i}(\psi_1+\psi_{-1})}{2} & \frac{2\psi_0}{\sqrt{6}} \end{pmatrix}. \label{eq:spin2tracelesstensor}
	\end{align} 
	If we use it, the problem reduces to that of $d$-wave superfluids \cite{PhysRevA.9.868}. The scalars are written as $ \rho=\operatorname{tr}T^\dagger T,\ \Theta=\operatorname{tr}T^2, $ and $ \boldsymbol{M}^2=2\operatorname{tr}[T,T^\dagger]^2 $. The singlet trio amplitude, which is used to characterize the cyclic state and label the degenerated nematic states\cite{Kobayashi09}, is written as $ A_{30}=\operatorname{tr}T^3 $. (Note, however, that the condition $ \boldsymbol{M}^2=|\Theta|^2=0 $ uniquely determines the cyclic state.)  Since $ \boldsymbol{M}^2=0 \leftrightarrow [T,T^\dagger]=0 \leftrightarrow [\operatorname{Re}T,\operatorname{Im}T]=0 \leftrightarrow $ ``$ T $ is diagonalizable by a rotation matrix'', the line $ \boldsymbol{M}^2=0 $ is characterized by $ D_2 $-symmetry, as in Fig.~\ref{fig:spin2region}. \\ 
	\indent One application of Fig.~\ref{fig:spin2region} is the direct determination of the spin-2 BEC phase diagram\cite{PhysRevA.61.033607} without solving the Gross-Pitaevskii equation. The phases of uniform condensates are determined by minimizing the two-body interaction energy $ c_1\boldsymbol{M}^2+c_2|\Theta|^2 $ with fixed $ \rho $. However, the minimization of this function in the triangle region is just the well-known linear optimization problem in the convex region, thus we immediately conclude that the minimum occurs at vertices, namely, the ground state is ferromagnetic, nematic, or cyclic. \\ 
	\indent \textit{Proof of Eq.~(\ref{eq:mainclaim})} --- Let us define the variables $ \gamma_0,\gamma_1,\dots,\gamma_{2F} $ by $ \psi_0=\gamma_0 $, $ \psi_{\pm 2m}=\frac{1}{\sqrt{2}}(\gamma_{2m}\pm\mathrm{i}\gamma_{2m+F}) $ for $  m=1,\dots,\lfloor\tfrac{F}{2}\rfloor $ and $ \psi_{\pm(2m-1)}=\frac{1}{\sqrt{2}}(\pm\gamma_{2m-1}+\mathrm{i}\gamma_{2m-1+F}) $ for $ m=1,\dots,\lfloor\tfrac{F+1}{2}\rfloor $. Then, we have $ \rho=\sum_{m=0}^{2F} |\gamma_m|^2,\ \Theta=\sum_{m=0}^{2F} \gamma_m^2 $, and $ M_z=\mathrm{i}\sum_{m=1}^Fm(\gamma_m^*\gamma_{m+F}-\gamma_{m+F}^*\gamma_m) $. From the Binet-Cauchy identity, the relation $ \rho^2-|\Theta|^2=\frac{1}{2}\sum_{m,n=0}^{2F}|\gamma_m^*\gamma_n-\gamma_n^*\gamma_m|^2 $ follows. Using them, we obtain
		\begin{align*}
		&\rho^2-|\Theta|^2-\frac{M_z^2}{F^2}=\sum_{m=1}^{2F}|\gamma_0^*\gamma_m-\gamma_m^*\gamma_0|^2\nonumber \\
		&+\frac{1}{2}\sum_{m,n=1}^F\left( 1-\frac{mn}{F^2} \right)\left\{ |\gamma_m^*\gamma_n-\gamma_n^*\gamma_m|^2\right.\nonumber \\
		&\left.\qquad+|\gamma_{m+F}^*\gamma_{n+F}-\gamma_{n+F}^*\gamma_{m+F}|^2+2|\gamma_m^*\gamma_{n+F}-\gamma_{n+F}^*\gamma_m|^2 \right\} \nonumber \\
		&+\frac{1}{2}\sum_{m,n=1}^F\frac{mn}{F^2}\left\{ |\gamma_n^*\gamma_m-\gamma_m^*\gamma_n-\gamma_{n+F}^*\gamma_{m+F}+\gamma_{m+F}^*\gamma_{n+F}|^2\right.\nonumber \\
		&\qquad\qquad \left.+|\gamma_{m+F}^*\gamma_n-\gamma_{n+F}^*\gamma_m-\gamma_m^*\gamma_{m+F}+\gamma_n^*\gamma_{n+F}|^2 \right\}\ge0.
	\end{align*}
	Since this inequality is proved without any constraint for $ \psi_m $'s, as a special subset of this proof, any spinor with $ M_x=M_y=0 $ also satisfies this inequality. Furthermore,  by $ SO(3) $-rotation, every spinor can be transformed to the one whose magnetization vector only has a $ z $-component. Thus, we conclude that the inequality  $ \rho^2-|\Theta|^2-\frac{\boldsymbol{M}^2}{F^2}\ge0 $ holds for all spinors. The equality occurs when (i) $ (\gamma_0,\dots,\gamma_{2F}) \propto (\gamma_0^*,\dots,\gamma_{2F}^*) $ or (ii)  $ \gamma_F $ and $ \gamma_{2F} $ are arbitrary and all other $ \gamma_m $'s vanish. The case (i) corresponds to TRS states $ \boldsymbol{\psi}\propto \mathcal{T}\boldsymbol{\psi} $. The case (ii), corresponding to $ C_{2Fv} $-symmetric states,  emerges because the term $ m=n=F $ in the second summand of the above expression does not contribute since $ 1-\frac{mn}{F^2}=0 $. $\blacksquare$ \\ 
	\indent In summary, we have shown the inequality (\ref{eq:mainclaim}) for spinor BECs valid for arbitrary integer spins. The symmetries of spinors in the triangle region Fig.~\ref{fig:spinFtriangle} are elucidated using MRs. The result is illustrated by the example of spin-2 BECs. The search for similar inequalities between other invariants, e.g., magnitude of nematic tensors and singlet trio amplitudes, is left to be open. The complicated phase diagram in spin-3 BECs \cite{PhysRevLett.96.190404,PhysRevLett.96.190405,PhysRevA.76.013605,PhysRevA.84.053616} suggests that the inequality including nematic tensors might not be simple. Inequalities for the systems with different group symmetries will be also worth investigating.
\begin{acknowledgment}
 The author is grateful to Muneto Nitta, Michikazu Kobayashi, and Shingo Kobayashi for fruitful discussions.
\end{acknowledgment}
\bibliographystyle{jpsj.bst}

\end{document}